\documentclass{kkzm12}

\usepackage[T1]{fontenc}
\usepackage[cp1250]{inputenc}

\usepackage{pdflscape}
\usepackage{graphicx}
\usepackage{algorithmic}
\usepackage{algorithm}
\usepackage{amssymb}
\usepackage{amsmath}
\usepackage{amsfonts}
\newcommand{\ud}{\mathrm{d}}
\newcommand{\E}[1]{\operatorname{E}\left[ #1 \right]}

\newcommand{\Cov}[1]{\operatorname{Var}\left[ #1 \right]}
\newcommand{\be}{\begin{equation}} \newcommand{\ee}{\end{equation}}
\newcommand{\bd}{\begin{displaymath}} \newcommand{\ed}{\end{displaymath}}

\title{The Laplace Motion in Phylogenetic Comparative Methods}
\author{Krzysztof Bartoszek}

\skrocone{K. Bartoszek}{Phylogenetic Laplace Motion} 

\affiliation{
\text{
Department of Mathematical Sciences,}
\text{Chalmers University of Technology and the University of Gothenburg} \\
\text{412 96, G\"oteborg Sweden,} \\
\text{\email{krzbar@chalmers.se}}\\
}

\begin{document}
\maketitle

\begin{abstract}
The majority of current phylogenetic comparative methods assume that the stochastic evolutionary
process is homogeneous over the phylogeny or offer relaxations of this in rather limited 
and usually parameter expensive ways. Here we make a preliminary investigation, by
means of a numerical experiment, whether the Laplace motion process can offer an alternative approach.
\end{abstract}

\section{Introduction -- Phylogenetic Comparative Methods}
It is by now well established that when doing a between species analysis 
(of some traits common to these species) one needs to take
into consideration that the data points (usually means of a number of individuals from each species)
could potentially come from a dependent sample. This dependence is due to the species'
shared evolutionary history. Due to phylogenetic inertia species which diverged
more recently are expected to have more similar trait values. This was noticed from 
the very birth of the theory of evolution \cite{CDar} but only recent availability
of computational power and genomic data, from which we can derive evolutionary
histories, allowed us to start developing and using phylogenetic comparative
methods in practice. The main challenge with comparative data 
is the type of sample we observe. We do not observe a time series trajectory
but we only (with some exceptions e.g. where fossil data is included, but in this situation one runs into dating issues) 
observe the trait values of currently alive species which contain a (better or worse estimated) 
branching structure behind them. This means that we do not have a natural way of exploiting
independent increments between observations.
\par
The first proposed model \cite{JFel1985} of continuous trait (labelled $X$) evolution 
was a Brownian motion model, $\ud X(t) =\sigma\ud B_{t}, X(0)=X_{0}$, where $B_{t}$ is
the Wiener process and $X_{0}$ is the trait value of the common ancestor of all of the studied species.
If we denote by $\mathbf{T}=[t_{a_{ij}}]_{1\le i,j \le n}$ the matrix of between species
divergence times ($t_{a_{ij}}$ is the time of divergence of species $i$ and $j$) then under this model
the mean and variance of our trait sample $\vec{x}=[x_{i}]_{1\le i \le n}$ are 
\mbox{
$\E{\vec{x}} = X_{0}\vec{1}_{n}$} and 
\mbox{
$\Cov{\vec{x}} = \sigma^{2}\mathbf{T}$}. This simple 
form makes the Brownian motion model tractable. Its major drawback however is that with time
the variance will be going to infinity and other models based on the Ornstein--Uhlenbeck
process have been studied \cites{JFel1988,THan1997,MButAKinOUCH,THanSOrz2005,THanJPieSOrzSLOUCH,ALabJPieTHan2009,KBarJPieSAndPMosTHan}.
One is naturally interested in estimating the process parameters and then discussing
to what biological properties they correspond to.
In the current study we will however not consider models more complicated than Brownian motion as we will be concentrating on another
aspect of phylogenetic comparative methods.

\section{Uncertainty in the phylogeny, relaxing process homogeneity}
The vast majority of programs for phylogenetic comparative analysis assume
that the provided phylogeny is completely resolved and are limited
in allowing process parameters to vary over different parts of the phylogenetic tree.
The current wealth of sequence data will provide us with better and better 
phylogenies but uncertainty is something that will never be completely removed.
The majority of current methods e.g. \cites{JParARamOPyb2008,MButAKinOUCH,Brownie} advocate to use an ensemble of (plausible)
phylogenies, on each one run the analysis and then weight the results (either uniformly or with some
prescribed weights to each phylogeny e.g. posterior probabilities from a Bayesian tree estimation
procedure). Alternatively one can run a joint MCMC that samples both the phylogeny and
stochastic process parameters that evolve on it \cites{JHueBRanJMas2000,MPagFLut2002,JHueBRan2003}.
These approaches are unfortunately computationally very demanding.
\par
One can also question the assumption of process homogeneity over the whole tree. It is 
of course possible to study models where the parameters differ over different parts
of the phylogeny \cites{MButAKinOUCH,Brownie,ALabJPieTHan2009}. This however causes an inflation in the 
number of parameters which results in more difficult and unreliable (due to small sample sizes)
estimation. One approach to this was suggested in \cite{PLemARamJWelMSuch2010} where it is 
proposed to rescale each branch length by a random variable from e.g. a gamma($\nu/2$,$\nu/2$)
or log--normal($1$,$\sigma$) distribution (see also \cite{ADruSHoMPhiARam2006}). This can be interpreted as that on each branch
the trait will have its unique speed of evolution but only at the cost of a single parameter describing the
distribution. This sort of time change can also be viewed as uncertainty in the phylogeny --- in the branch length estimates. 
Another possibility is presented in \cite{JEasetal2011}, where a Bayesian model selection procedure searches for Brownian motion rate
changes over the phylogeny.

\section{Laplace Motion}
The Laplace motion (also known as the variance gamma process) 
can be represented as a Brownian motion with drift with a
random time change given by a gamma process \cite{DMadPCarECha1998}.
Following \cite{DMadPCarECha1998} let $B_{\theta,\sigma}(t)$ be a Brownian motion with drift,
\be
B_{\theta,\sigma}(t) = \theta t + \sigma B(t),
\ee
and then we define, again after \cites{DMadPCarECha1998,KBogKPodIRych2010,TKozKPodIRych2010}, 
the gamma process $\gamma_{\nu,\beta}(t)$ as the process
of independent gamma increments over non--overlapping time intervals. The density of the increment 
$g = \gamma_{\nu,\beta}(t+h)-\gamma_{\nu,\beta}(t)$ over 
a time interval of length $h$ is given by the gamma density with shape parameter $h/\nu$ and scale parameter $\beta$,
\be
f_{h ; \nu,\beta}(g) = 
\left(\Gamma\left(\frac{h}{\nu} \right)\right)^{-1} \beta^{\frac{h}{\nu}} g^{\frac{h}{\nu}-1}e^{-\beta g}
\ee
Now we define the Laplace motion as,
\be 
X_{\theta,\sigma,\nu}(t) = X_{0}+B_{\theta,\sigma}(\gamma_{\nu,\nu}(t)).
\ee
In our particular application we consider the Laplace motion 
with no drift, $\theta \equiv 0$, also known as the symmetric
variance gamma process \cites{DMadESen1990,DMadPCarECha1998}, for which given a fixed $t$
the following equality in distribution holds,
\be \label{eqLapBM}
X_{\sigma,\nu}(t) = B_{\sigma}(\gamma_{\nu,\nu}(t)) \stackrel{\mathcal{D}}{=} X_{0}+\sqrt{\gamma_{\nu,\nu}(t)} \sigma B(t).
\ee
Naturally a non--zero drift
has biological interest however at this point we do not consider it.
The idea of using a Laplace motion is closely related to the suggestion of \cite{PLemARamJWelMSuch2010},
but it is put in a formal mathematical framework of L\'evy processes.

\section{Laplace Motion estimation}
To estimate the model parameters of interest, $X_{0}$, $\sigma^{2}$ and $\nu$ we use
an EM algorithm, \mbox{following \cite{TKozKPodIRych2010},} which exploits the representation in Eq. \eqref{eqLapBM} 
of the Laplace motion by treating the unobservable gamma random variables for the branch lengths (variances) as 
missing values. We also exploit that we only have observations of the tip species. Due to this, 
from the perspective of a phylogenetic comparative method, a Laplace motion is equivalent
to changing each branch $i$ of length $t_{i}$ to a gamma random variable
with mean $t_{i}$ and variance $\nu t_{i}$. Notice that in this parametrization
of a gamma random variable, $\sigma^{2}$ is not a superfluous parameter as it will not
``disappear'' into the gamma process' scale parameter due to us forcing a relationship
between the shape and scale. 
Multiplying a gamma random variable by a constant is equivalent to multiplying the scale
by the same value but it does not relate to the shape. In the limit as $\nu \rightarrow 0$
we arrive at the original tree. The main computational challenge is how to effectively 
do the expectation (E) step as we do not observe a time series process. The main 
aim of this work is an initial consideration of introducing the Laplace motion so we 
made do with an approximate numerical treatment of the problem.
For the maximization step we employ a simulated annealing
type of search. 
We used \mbox{R 2.15 \cite{R}} (on an Intel Core i5 $2.3$GHz machine running Suse Linux 12.1) 
to implement the estimation procedure described in Alg. \ref{algEMBMLap}.
In it we allow for a non--homogeneous Laplace motion, i.e. we allow the parameter
$\nu$ describing it to vary over predefined by the user branches of the phylogeny (regimes).
A user is free to specify a specific value for each branch but then we run into obvious 
estimation problems. The natural approach is to group branches ($\nu$s) according 
to some shared trait/environmental variable (see the biological example where we have
one $\nu$ for each breeding group size).

\begin{algorithm}[!ht]
\caption{Approximate EM algorithm for MLE of phylogenetic Laplace motion}\label{algEMBMLap}
\begin{algorithmic}
\STATE $i:=1$, $\mathcal{L}_{0}= -\infty$, $\nu_{1}=$starting value, $\tau_{0}=50$
\WHILE{
$i\le 60$ \textbf{and} $\vert \mathcal{L}_{i} - \mathcal{L}_{i-1}  \vert > \epsilon$
}
    \IF{previous iteration was rejected}
	\STATE $\mathcal{L}_{i}= \mathcal{L}_{i-1}$
    \ENDIF
    \FOR{$k=1$ to $1000$}
	\STATE $Tree_{k} := Original~Tree$, $\mathcal{L}_{T_{k}}=0$
	\FOR{
	    each branch $j$ in $Tree_{k}$}
		\STATE $t_{j} = \Gamma(shape=T_{j}/\nu_{i}[j],scale=\nu_{i}[j])$ \COMMENT{
				$T_{j}$ is branch $j$'s length in the original tree}
		\STATE $\mathcal{L}_{T_{k}}+= \log(f_{T_{j},\nu}(t_{j}))$ \COMMENT{
					    $f_{T_{j},\nu}(\cdot)$ is the appropriate gamma density}
	\ENDFOR
	\STATE $(\widehat{X_{0}}_{i_{k}},\widehat{\sigma^{2}}_{i_{k}})=$ MLE conditional on $Tree_{k}$ and trait data vector $\vec{x}$ under the Brownian motion model by mvSLOUCH \cite{KBarJPieSAndPMosTHan}
	\STATE $\mathcal{L}_{x_{k}}=$log--likelihood of $(\widehat{X_{0}}_{i_{k}},\widehat{\sigma^{2}}_{i_{k}})$ conditional on $\vec{x}$ and $Tree_{k}$
    \ENDFOR
    \STATE $\mathcal{L}_{T}=\sum\limits_{k}\mathcal{L}_{T_{k}}$ \COMMENT{The following four are heuristic formulae that were found to work well}
    \STATE $\mathcal{L}_{i}=\sum\limits_{k}\left(\mathcal{L}_{T_{k}}+ \mathcal{L}_{x_{k}} - \mathcal{L}_{T}\right)$
    \STATE $\widehat{X_{0}}_{i}=\sum\limits_{k}\widehat{X_{0}}_{i_{k}}\cdot \exp(\mathcal{L}_{T_{k}}+\mathcal{L}_{x_{k}}-\mathcal{L}_{T}-\mathcal{L}_{i})$
    \STATE $\widehat{\sigma^{2}}_{i}=\sum\limits_{k}\widehat{\sigma^{2}}_{i_{k}}\cdot \exp(\mathcal{L}_{T_{k}}+\mathcal{L}_{x_{k}}-\mathcal{L}_{T}-\mathcal{L}_{i})$
    \IF{
	$\mathcal{L}_{i} \le \mathcal{L}_{i-1}$ \textbf{and} $U>\exp((\mathcal{L}_{i}- \mathcal{L}_{i-1})/\tau_{i}) $
	 \COMMENT{$U$ is uniform on $[0,1]$} }
	\STATE $\widehat{X_{0}}_{i}= \widehat{X_{0}}_{i-1}$, $\widehat{\sigma^{2}}_{i}=\widehat{\sigma^{2}}_{i-1}$ \COMMENT{do not update}
    \ENDIF
    \STATE $\nu_{i+1}=\exp(\log(\nu_{i})+\zeta\vert Z\vert)$ \COMMENT{$Z$ is normal with mean $0$ and variance dependent on $\nu_{i}$, \mbox{$\zeta\in \{-1,1\}$} depending which direction seems better locally}
    \STATE $\tau_{i+1}=\tau_{0}/i$
    \STATE $i++$    
\ENDWHILE
\end{algorithmic}
\end{algorithm}

\section{Data example Cervidae female body size}
\begin{figure}[!ht]
\centering
\includegraphics[height=7cm]{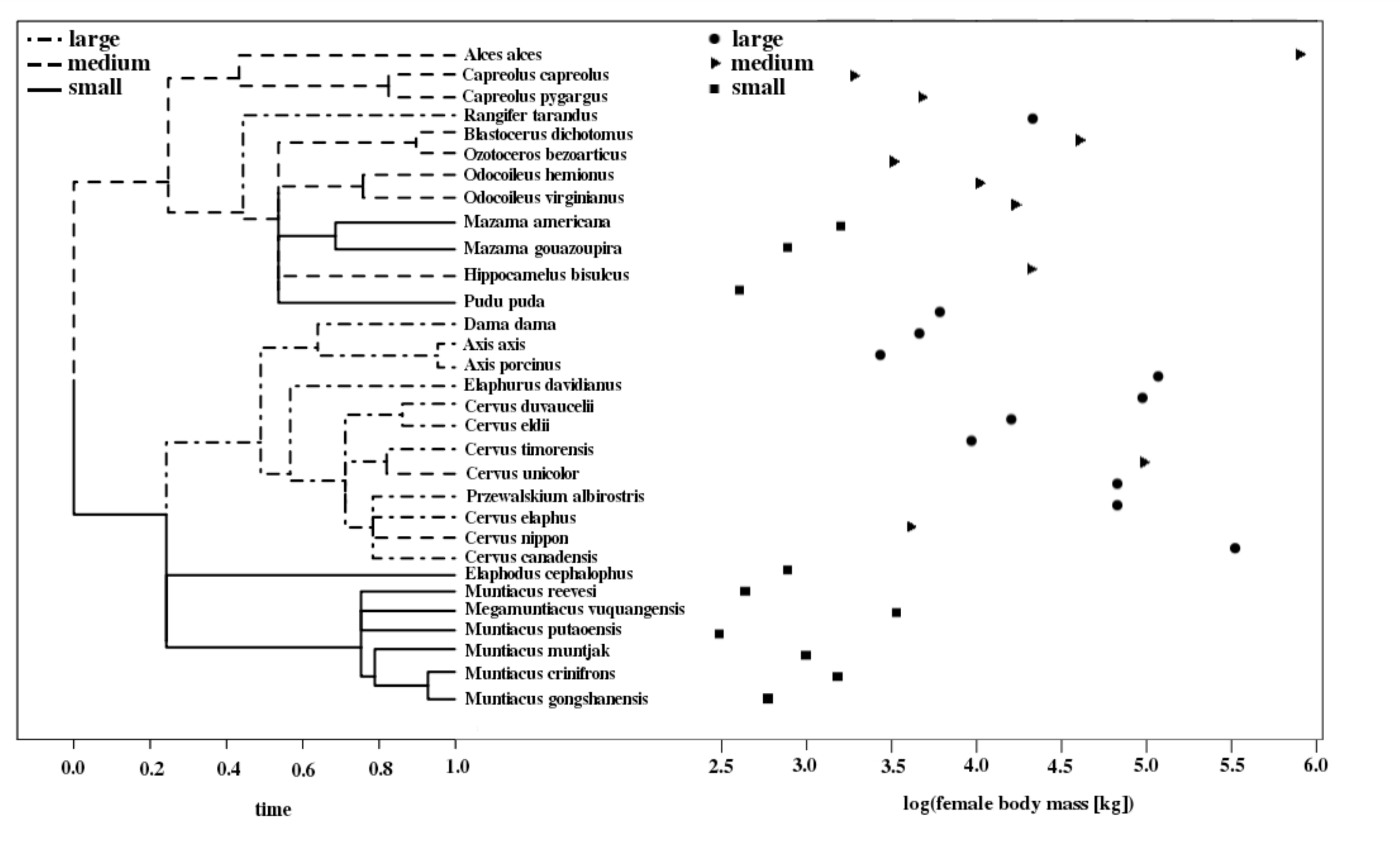}
\caption{Left: Cervidae phylogeny \cite{MFerEVrb2004}, right: logarithm of female body size and breeding group size.}
\label{figDeerFBM}
\end{figure}

As a proof of concept example we re--analyze a portion of the Cervidae data set of
\cite{FPlaetal2011}. The data set consists of averages of measurements of male antler length (mm),
male body mass (kg), female body mass (kg) in $31$ deer species along with their phylogeny derived from
\cite{MFerEVrb2004}. 
In addition the breeding 
group size (discrete variable with three levels: $1$--$2$, $3$--$5$ and $>5$) of each species is recorded
and also the mating tactic (discrete variable with four levels: harem, territorial, tending and monogamous) 
is recorded 
(missing in \emph{Megamuntiacus vuquangensis}). The aim of \cite{FPlaetal2011} was to study how
male antler length depends on the remaining measured variables. In 
\cite{KBarJPieSAndPMosTHan} we were able to improve on this analysis
due to our newly developed mvSLOUCH R software. We found that both 
male antler length and male body mass were jointly responding to changes
in female body mass and breeding group size. The logarithm of female body mass was 
found to be best explained (AIC$_{c}$) by a Brownian motion. The results
and biological conclusions are consistent with those of \cite{FPlaetal2011} except 
that a more refined model provides better explanation for them. In this study
we consider just the female body mass and see whether it can be better explained by
a Brownian motion with different rates for different breeding group sizes
or by a Laplace motion with one $\nu$ parameter or a different one for each 
breeding group size. We also checked whether the diffusion parameter of the 
Brownian motion aids estimation in the Laplace motion models. 
The female body size and deer phylogeny is presented in Fig. \ref{figDeerFBM}.
The estimation results are presented in Tabs. \eqref{tbModelComp} and \eqref{tbModelEst}.

\begin{table}[!ht]
\begin{center}
\caption{Summary of Cervidae model comparisons. Model marked with 
$^{\ast}$ was analyzed with Brownie \cite{Brownie}, remaining by a front--end to mvSLOUCH.
In Fig. \ref{figmodels} we can see the relationships between the models.}\label{tbModelComp}
\begin{tabular}{|c|c|c|c|c|}
\hline
Model & df & Log--likelihood & AIC & AIC$_{c}$ \\
\hline
BM & $2$ & $-36.140$ & $76.279$ & $76.708$ \\
\hline
BM + BGS$^{\ast}$ & $4$ & $-34.815$ & $77.629$ & $79.168$ \\
\hline
\textbf{Laplace} & $3$ & $-34.798$ & $\mathbf{75.596}$ & $\mathbf{76.484}$ \\
\hline 
Laplace + BGS & $5$ & $-34.741$ & $79.481$ & $81.881$\\
\hline
Laplace ($\sigma^{2}=1$) & $2$ & $-37.017$ & $78.034$ & $78.463$ \\
\hline
 Laplace + BGS ($\sigma^{2}=1$) & $4$ & $-36.945$ & $81.890$ & $83.429$\\
\hline
\end{tabular}
\end{center}
\end{table}

\begin{table}[!ht]
\begin{center}
\caption{Summary of Cervidae estimation. Values marked with 
$^{\ast}$ were estimated by Brownie \cite{Brownie}, remaining by a front--end to mvSLOUCH.
Notice that $\sigma^{2}$ is a relative value as we set the tree's height as $1$.}\label{tbModelEst}
\begin{tabular}{|c|c|c|c|c|c|c|}
\hline
Parameter & BM & BM+BGS & \textbf{Laplace} & Laplace + BGS & Laplace & Laplace + BGS \\
\hline
&&&&&($\sigma^{2}=1$) & ($\sigma^{2}=1$) \\
\hline
$X_{0}$ & $3.835$ & $3.459^{\ast}$ & $\mathbf{3.838}$ & $3.878$ & $3.767$ & $3.792$\\
\hline
$\sigma^{2}$ & $1.396$ & --- & $\mathbf{1.379}$ & $1.269$ &  --- & --- \\
\hline
$\sigma^{2}_{small}$ & --- & $0.642^{\ast}$ & --- &--- &--- &--- \\
\hline
$\sigma^{2}_{medium}$ & --- & $2.568^{\ast}$ &  --- &--- &--- &--- \\
\hline
$\sigma^{2}_{large}$ & --- &  $1.251^{\ast}$ & --- &--- &--- &--- \\
\hline
$\nu$ & --- & --- & $\mathbf{0.897}$ & ---& $0.604$ & --- \\
\hline
$\nu_{small}$ & --- & --- & --- & $0.712$ & --- & $0.248$\\
\hline
$\nu_{medium}$ & --- & --- & --- & $1.006$ & --- & $0.100$\\
\hline
$\nu_{large}$ & --- & --- & --- & $0.816$ & --- & $0.049$\\
\hline
\end{tabular}
\end{center}
\end{table}

\begin{figure}[!ht]
\centering
\includegraphics[height=4cm]{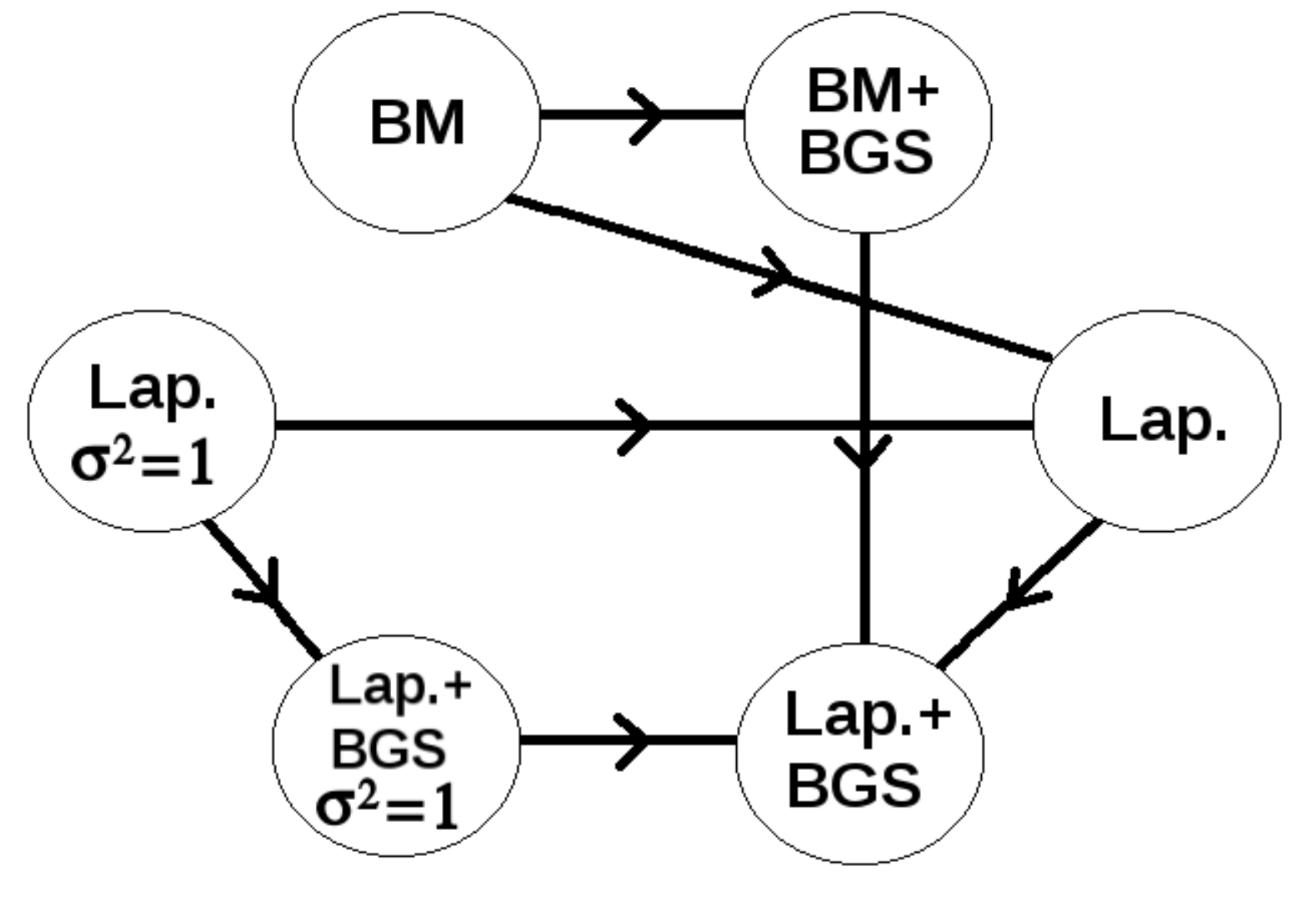}
\caption{Relationships between models, an arrow from model $i$ to $j$ indicates that model $i$ is a direct submodel of $j$
(we do not indicate transitive relationships).}
\label{figmodels}
\end{figure}
\par
Using the AIC$_{c}$ to choose between competing models we can see that, from the set of candidate models,
the homogeneous Laplace motion with $\sigma^{2}$ unknown best explains the female body size data. 
This suggests that the rate of trait evolution is not homogeneous but currently available methods would
not detect this as they would have to be overly parametrized. 

\section{Conclusions and future work}
The aim of this work was to describe how the Laplace motion model would fit in the phylogenetic comparative
methods field and whether it could be a potential alternative to models with different diffusion coefficients
on different parts of the phylogeny. Such models are expensive in terms of parameters and with commonly small samples could be rejected in 
favour of a homogeneous model. The Laplace motion allows one to relax the process
homogeneity assumption without paying too much in parameters. Such a situation was seen in
the presented here Cervidae data example. This work is however only a preliminary investigation to gain
intuitions for the model. A detailed analytical study is required in order to achieve a stabler and more
formal estimation procedure, find its estimability properties 
and be able to couple the variance gamma process with Ornstein--Uhlenbeck 
type evolutionary processes.

\section{Acknowledgments}
The author thanks Joachim Domsta, Krzysztof Podg\'orski and Igor Rychlik for helpful discussions and comments.
K.B. was supported by the Centre for Theoretical Biology at the University of Gothenburg, 
Stiftelsen f\"or Vetenskaplig Forskning och Utbildning i Matematik
(Foundation for Scientific Research and Education in Mathematics), 
Knut and Alice Wallenbergs travel fund, Paul and Marie Berghaus fund, the Royal Swedish Academy of Sciences,
and Wilhelm and Martina Lundgrens research fund.

\begin{bibliography}
\vspace*{-0.3cm}
  \bib{KBarJPieSAndPMosTHan}{article}{
    author={K. Bartoszek},
    author={J. Pienaar},
    author={P. Mostad},
    author={S. Andersson},
    author={T. F. Hansen},
    title={A comparative method for studying multivariate adaptation},
    journal={},
    volume={},
    date={},
    pages={working paper},
  }

  \bib{KBogKPodIRych2010}{article}{
    author={K. Bogsj\"o},
    author={K. Podg\'orski},
    author={I. Rychlik},
    title={Models for road surface roughness},
    journal={Preprint Matematiska vetenskaper G\"oteborg},
    volume={2010:42},
    date={2010},
    pages={},
  }

  \bib{MButAKinOUCH}{article}{
    author={M. A. Butler},
    author={A. A. King},
    title={Phylogenetic comparative analysis: a modelling approach for adaptive evolution},
    journal={Am. Nat.},
    volume={164},
    date={2004},
    pages={683-695},
  }

  \bib{CDar}{book}{
    author={C. Darwin},
    title={On the Origin of Species},
    publisher={John Murray},
    place={London},
    date={1859},
  }
  
  \bib{ADruSHoMPhiARam2006}{article}{
    author={A. J. Drummond},
    author={S. Y. W. Ho},
    author={M. J. Phillips},
    author={A. Rambaut},
    title={Relaxed Phylogenetics and Dating with Confidence},
    journal={PLoS Biol.},
    volume={4},
    date={2006},
    pages={699-710},
  }

  \bib{JEasetal2011}{article}{
    author={J. M. Eastman},
    author={M. E. Alfaro},
    author={P. Joyce},
    author={A. L. Hipp},
    author={L. J. Harmon},
    title={A novel comparative method for identifying shifts in the rate of character evolution on trees},
    journal={Evolution},
    volume={65},
    date={2011},
    pages={3578-3589},
  }

\bib{JFel1985}{article}{
    author={J. Felsenstein},
    title={Phylogenies and the comparative method},
    journal={Am. Nat.},
    volume={125},
    date={1985},
    pages={1-15},
  }

  \bib{JFel1988}{article}{
    author={J. Felsenstein},
    title={Phylogenies and Quantitative Characters},
    journal={Annu. Rev. Ecol. Syst.},
    volume={19},
    date={445-471},
    pages={1988},
  }

  \bib{MFerEVrb2004}{article}{
    author={M. H. Fern\'andez},
    author={E. S. Vrba},
    title={A complete estimate of the phylogenetic relationships in \textnormal{{R}uminantia}: a dated species--level supertree of the extant ruminants},
    journal={Biol. Rev.},
    volume={80},
    date={2004},
    pages={269-302},
  }

  \bib{THan1997}{article}{
    author={T. F. Hansen},
    title={Stabilizing selection and the comparative analysis of adaptation},
    journal={Evolution},
    volume={51},
    date={1997},
    pages={1341-1351},
  }

  \bib{THanSOrz2005}{article}{
    author={T. F. Hansen},
    author={S. H. Orzack},
    title={Assessing current adaptation and phylogenetic inertia as explanations of trait evolution: the need for controlled comparisons},
    journal={Evolution},
    volume={59},
    date={2005},
    pages={2063-2072},
  }
 
  \bib{THanJPieSOrzSLOUCH}{article}{
    author={T. F. Hansen},
    author={J. Pienaar},
    author={S. H. Orzack},
    title={A comparative method for studying adaptation to a randomly evolving environment},
    journal={Evolution},
    volume={62},
    date={2008},
    pages={1965-1977},
  }

  \bib{JHueBRanJMas2000}{article}{
    author={J. P. Huelsenbeck},
    author={B. Rannala},
    author={J. P. Masly},
    title={Accommodating Phylogenetic Uncertainty in Evolutionary Studies},
    journal={Science},
    volume={288},
    date={2000},
    pages={2349-2350},
  }

  \bib{JHueBRan2003}{article}{
    author={J. P. Huelsenbeck},
    author={B. Rannala},
    title={Detecting correlation between characters in a comparative analysis with uncertain phylogeny},
    journal={Evolution},
    volume={57},
    date={2003},
    pages={1237-1247},
  }

  \bib{TKozKPodIRych2010}{article}{
    author={T. J. Kozubowski},
    author={K. Podg\'orski},
    author={I. Rychlik},
    title={Multivariate Generalized Laplace Distributions and Related Random Fields},
    journal={Preprint Matematiska vetenskaper G\"oteborg},
    volume={2010:47},
    date={2010},
    pages={},
  }

  \bib{ALabJPieTHan2009}{article}{
    author={A. Labra},
    author={J. Pienaar},
    author={T. F. Hansen},
    title={Evolution of Thermal Physiology in \textit{Liolaemus} Lizards: Adaptation, Phylogenetic Inertia, and Niche Tracking},
    journal={Am. Nat.},
    volume={174},
    date={2009},
    pages={204-220},
  }

  \bib{PLemARamJWelMSuch2010}{article}{
    author={P. Lemey},
    author={A. Rambaut},
    author={J. J. Welch},
    author={M. A. Suchard},
    title={Phylogeography takes a relaxed random walk in continuous space and time},
    journal={Mol. Biol. Evol.},
    volume={27},
    date={2010},
    pages={1877-1885},
  }

  \bib{DMadPCarECha1998}{article}{
    author={D. B. Madan},
    author={P. P. Carr},
    author={E. C. Chang},
    title={The Variance Gamma Process and Option Pricing},
    journal={Eur. Financ. Rev.},
    volume={2},
    date={1998},
    pages={79-105},
  }

  \bib{DMadESen1990}{article}{
    author={D. B. Madan},
    author={E. Seneta},
    title={The Variance Gamma (V.G.) Model for Share Market Returns},
    journal={J. Bus.},
    volume={63},
    date={1990},
    pages={511-524},
  }

  \bib{Brownie}{article}{
    author={B. C. O'Meara},
    author={C. An\'e},
    author={M. J. Sanderson},
    author={P. C. Wainwright},
    title={Testing for different rates of continuous trait evolution using likelihood},
    journal={Evolution},
    volume={60},
    date={2006},
    pages={922-933},
  }

  \bib{MPagFLut2002}{article}{
    author={M. Pagel},
    author={F. Lutzoni},
    title={Accounting for phylogenetic uncertainty in comparative studies of evolution and adaptation},
    journal={Biological Evolution and Statistical Physics},
    volume={},
    date={2002},
    pages={146-161},
  }

  \bib{JParARamOPyb2008}{article}{
    author={J. Parker},
    author={A. Rambaut},
    author={O. G. Pybus},
    title={Correlating viral phenotypes with phylogeny: {A}ccounting for phylogenetic uncertainty},
    journal={Infect. Genet. Evol.},
    volume={8},
    date={2008},
    pages={239-246},
  }

  \bib{FPlaetal2011}{article}{
    author={F. Plard},
    author={C. Bonenfant},
    author={J. M. Gaillard},
    title={Revisiting the allometry of antlers among deer species: male--male sexual competition as a driver},
    journal={Oikos},
    volume={120},
    date={2011},
    pages={601-606},
  }
                        
  \bib{R}{book}{
    author={{R Development Core Team}},
    title={{R: A Language and Environment for Statistical Computing}},
    publisher={{R Foundation for Statistical Computing}},
    place={Vienna},
    date={2010},
  }

\end{bibliography}

\end{document}